\documentclass[prl,twocolumn,superscriptaddress,amsmath,amssymb]{revtex4}
\usepackage{times}
\usepackage{graphicx}
\usepackage{amsmath,amssymb}
\usepackage{lipsum, babel}
\usepackage{afterpage}
\usepackage{amsfonts}
\usepackage{amssymb}
\usepackage{graphicx}
\usepackage{hyperref}
\usepackage{mathtools}
\usepackage{verbatim}
\usepackage{epstopdf}
\usepackage{subfigure}
\usepackage{latexsym}
\usepackage{dcolumn}
\usepackage{epsf}
\usepackage{float}
\usepackage[table]{xcolor}
\usepackage{multirow}
\usepackage{makecell}

\begin{document}

\preprint{APS/123-QED}
	 
\title{Why are there six degrees of separation in a social network?}

\author{I. Samoylenko}
\thanks{These Authors equally contributed to the Manuscript}
\affiliation{\noindent \textit{Moscow Institute of Physics and Technology, Dolgoprudny, Moscow Region, 141701, Russia}}
\affiliation{\noindent \textit{National Research University Higher School of Economics, 6 Usacheva str., Moscow 119048, Russia}}

\author{D. Aleja}
\thanks{These Authors equally contributed to the Manuscript}
\affiliation{\noindent \textit{Universidad Rey Juan Carlos, Calle Tulip\'an s/n, 28933, M\'ostoles, Madrid, Spain}}
\affiliation{\noindent \textit{Department of Internal Medicine, University of Michigan, Ann Arbor, MI, USA}}

\author{E. Primo}
\affiliation{\noindent \textit{Universidad Rey Juan Carlos, Calle Tulip\'an s/n, 28933, M\'ostoles, Madrid, Spain}}

\author{K. Alfaro-Bittner}
\affiliation{\noindent \textit{Universidad Rey Juan Carlos, Calle Tulip\'an s/n, 28933, M\'ostoles, Madrid, Spain}}

\author{E. Vasilyeva}
\affiliation{\noindent \textit{Moscow Institute of Physics and Technology, Dolgoprudny, Moscow Region, 141701, Russia}}
\affiliation{\noindent \textit{P.N. Lebedev Physical Institute of the Russian Academy of Sciences, 53 Leninsky prosp., 119991 Moscow, Russia}}

\author{K. Kovalenko}
\affiliation{\noindent \textit{Scuola Superiore Meridionale, School for Advanced Studies, Naples, Italy}}

\author{D. Musatov}
\affiliation{\noindent \textit{Moscow Institute of Physics and Technology, Dolgoprudny, Moscow Region, 141701, Russia}}
\affiliation{\noindent \textit{Russian Academy of National Economy and Public Administration, pr. Vernadskogo, 84, Moscow, 119606, Russia}}

\author{A.M. Raigorodskii}
\affiliation{\noindent \textit{Moscow Institute of Physics and Technology, Dolgoprudny, Moscow Region, 141701, Russia}}
\affiliation{\noindent \textit{Moscow State University, Leninskie Gory, 1, Moscow, 119991, Russia}}

\author{R. Criado}
\affiliation{\noindent \textit{Universidad Rey Juan Carlos, Calle Tulip\'an s/n, 28933, M\'ostoles, Madrid, Spain}}

\author{M. Romance}
\affiliation{\noindent \textit{Universidad Rey Juan Carlos, Calle Tulip\'an s/n, 28933, M\'ostoles, Madrid, Spain}}

\author{D. Papo}
\affiliation{\noindent \textit{Department of Neuroscience and Rehabilitation, University of Ferrara, Ferrara, Italy}}

\author{M. Perc}
\affiliation{\noindent \textit{Faculty of Natural Sciences and Mathematics, University of Maribor, Koro{\v s}ka cesta 160, 2000 Maribor, Slovenia}}
\affiliation{\noindent \textit{Department of Medical Research, China Medical University Hospital, China Medical University, Taichung 404332, Taiwan}}
\affiliation{\noindent \textit{Complexity Science Hub Vienna, Josefst{\"a}dterstra{\ss}e 39, 1080 Vienna, Austria}}
\affiliation{\noindent \textit{Alma Mater Europaea, Slovenska ulica 17, 2000 Maribor, Slovenia}}
\affiliation{\noindent \textit{Department of Physics, Kyung Hee University, 26 Kyungheedae-ro, Dongdaemun-gu, Seoul, Republic of Korea}}

\author{B. Barzel}
\affiliation{\noindent \textit{Department of Mathematics, Bar-Ilan University, Ramat-Gan, Israel}}
\affiliation{\noindent \textit{The Gonda Multidisciplinary Brain Research Center, Bar-Ilan University, Ramat-Gan, Israel}}
\affiliation{\noindent \textit{Network Science Institute, Northeastern University, Boston, MA., US}}

\author{S. Boccaletti}
\affiliation{\noindent \textit{Moscow Institute of Physics and Technology, Dolgoprudny, Moscow Region, 141701, Russia}}
\affiliation{\noindent \textit{Universidad Rey Juan Carlos, Calle Tulip\'an s/n, 28933, M\'ostoles, Madrid, Spain}}
\affiliation{\noindent \textit{CNR - Institute of Complex Systems, Via Madonna del Piano 10, I-50019 Sesto Fiorentino, Italy}}
\affiliation{\noindent \textit{Complex Systems Lab, Department of Physics, Indian Institute of Technology, Indore - Simrol, Indore 453552, India}}

\date{\today}

\begin{abstract}
	A wealth of evidence shows that real world networks are endowed with the small-world property i.e., that the maximal distance between any two of their nodes scales logarithmically rather than linearly with their size. In addition, most social networks are organized so that no individual is more than six connections apart from any other, an empirical regularity known as {\it the six degrees of separation}. Why social networks have this {\it ultra-small world} organization, whereby the graph’s diameter is independent of the network size over several orders of magnitude, is still unknown. We show that the `six degrees of separation' are the property featured by the equilibrium state of any network where individuals weigh between their aspiration to improve their centrality and the costs incurred in forming and maintaining connections. We show, moreover, that the emergence of such a regularity is compatible with all other features, such as clustering and scale-freeness, that normally characterize the structure of social networks.
	Thus, our results show how simple evolutionary rules of the kind traditionally associated with human cooperation and altruism can also account for the emergence of one of the most intriguing attributes of social networks.
\end{abstract}

\maketitle
In the short story "Chains" (1929), the Hungarian writer Frigyes Karinthy described a game where a group of people was discussing how the members
of the human society were closer together than ever before. To prove this point, one participant proposes that any person out of the entire
Earth population (around 1.8 billion at that time) could be reached using nothing except each personal network
of acquaintances, betting that the resulting chain would be of no more than five individuals.
The story coined the expression `six degrees of separation' to reflect the idea that all people of the world are
six or fewer social connections apart from each other. The concept was later generalized to that of `small world' networks,
where the maximal social distance (the diameter of the network) scales
logarithmically, rather than linearly, with the size of the population~\cite{watts}.

After early studies on the structure of social networks by Michael Gurevich~\cite{gure} and Manfred Kochen~\cite{koch}, Stanley Milgram performed his 1967 famous set of experiments on social distancing~\cite{milgram,milgram2}  where, with a limited sample of a thousand individuals, it was shown that people in the United States are indeed connected by a small number of acquaintances. Later on, Duncan Watts recreated Milgram's experiments with Internet email users~\cite{dodds} by tracking 24,163 chains aimed at 18 targets from 13 countries and confirmed that the average number of steps in the chains was around six. Furthermore, many experiments conducted at a planetary scale on various social networks verified the ubiquitous character of this feature: {\it i)} a 2007 study by Jure Leskovec and Eric Horvitz (with a data set of 30 billion conversations
among 240 million Microsoft Messenger users) revealed  the average path length to be 6~\cite{lesko,lesko2}, {\it ii)} the average degree of separation between two randomly selected Twitter users was found to be 3.435~\cite{bakhsh}, and {\it iii)} the Facebook's  network in 2011 (721 million users with 69 billion friendship links) displayed an average distance between nodes of 4.74~\cite{facebook}.

Such abundant and consistent evidence points to the fact that the structure of these networks radically differs from either that of regular networks (where the diameter scales linearly with the size) and that of classical small-world networks (where, instead, the scaling law is logarithmic) \cite{watts}.
A clear explanation of the mechanisms through which social networks organize into {\it ultra-small world} states (where the diameter does not depend on the system size over several orders of magnitude) is, however, still missing. Why does such a collective property emerge? What are its fundamental mechanisms?
Why is the common shortest path length between units of a social network six, rather than five or seven or any other number, implying an average distance which is also not far from six?

We here answer these questions in exact terms, by adopting a game theoretical approach for describing the network evolution, a line of studies which started almost five decades ago by Myerson~\cite{myerson1977graphs} analyzing cooperation structures in a wide class of games. A couple of decades later, games on adaptive networks were introduced, for instance by Jackson and Wolinsky~\cite{jackson1996strategic}, with the purpose of studying the stability and efficiency of social and economic networks when self-interested individuals could form or sever links with others. Further on, the influential work by Nowak and May~\cite{nowak_n92b} showed how spatial structure could provide an evolutionary escape hatch for cooperation in social dilemmas. Co-evolutionary networks have then been considered in a series of works where players could improve their topological position, for example by cutting links to defectors or rewiring their links to gain larger payoffs~\cite{ebel2002coevolutionary, zimmermann_pre04, zimmermann_pre05, pacheco_prl06, gomez-gardenes_prl07, gross_jrsi08}. Related research also covered game theoretical models as the basis for cooperation on networks~\cite{ohtsuki_n06}, for network formation and growth~\cite{goyal2005network, bloch2007formation, poncela_njp09, bei2011bounded}, as well as for agents to achieve a position of high centrality while minimizing the number of contacts they have to maintain~\cite{holme_prl06}.

So far, the few available studies on  ultra-small world states have focused on finding the relationship between the scaling properties of distances in a graph and those of the node's degree distribution. It was indeed proved that scale-free networks with degree distribution $p(k) \sim k^{-\gamma}$ and $2< \gamma <3$ (as it is observed in all real-world networks) display a scaling of the diameter as $D \sim \ln \ln N$ \cite{cohen2003}, which departs from the classic logarithmic scaling of small-world networks and yet maintains an explicit dependence on the network size $N$. On the other hand, scale-free networks featuring an asymptotically invariant shortest path (called {\it Mandala networks} \cite{mandala}) may be synthesized, which however have an associated value of $\gamma$ strictly equal to 2 and therefore do not match any case observed in real world.

Rather than being dependent on global (i.e., degree distribution) scaling properties, in this Article
we show that the mechanism behind such observed regularity can be found, instead, in a dynamic evolution of the network.
Precisely, we rigourously show that, when a simple compensation rule between the cost incurred by nodes in maintaining
connections and the benefit accrued by the chosen links is governing the evolution of a network, the asymptotic equilibrium state
(a Nash equilibrium where no further actions would produce more benefit than cost~\cite{nash_pnas50}), features a diameter
{\it which does not depend} on the system's size, and is equal to 6.
In other words, we theorematically prove that any network where nodes strive to increase their centrality by forming connections if and only if their cost is smaller than the payoff tends to evolve into an ultra-small world state endowed with the 'six degree of separation' property, irrespective of its initial structure.
Our study points out, therefore, that evolutionary rules of the kind traditionally associated with human cooperation and altruism~\cite{novak,rand1,rand2,fu1,fu2} can in fact account also for the emergence of this attribute of social networks. Furthermore, we show that such a global network feature can emerge even from situations where individuals have access to only partial information on the overall structure of connections, which is indeed the case in almost all social networks.

\section{Results}

\subsection{A game theoretical model for network evolution}

\begin{figure}[t!]
	\includegraphics[width=1\linewidth]{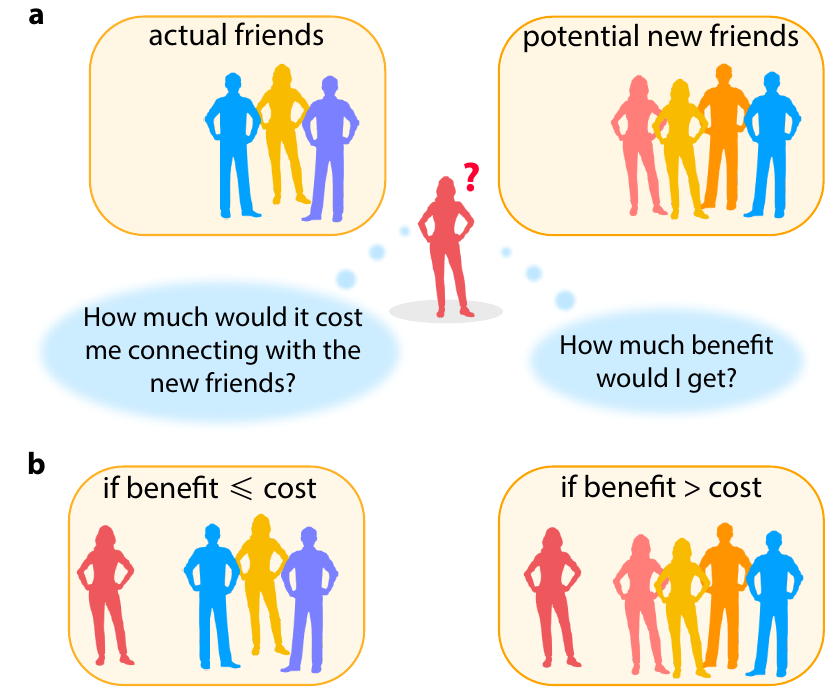}
	\caption{{\bf The game theoretical framework.} The structure of a social network evolves following simple rules of a game. Panel a: At each step of the game, the individuals forming part of the network (like the red woman in the picture) have to decide whether to stay with the neighborhood formed by their actual friends, or to change to another neighborhood formed by potential new friends. The current and new neighborhoods may overlap (in our picture, the blue man and the yellow woman are members of both sets). The decision is based on a careful evaluation of the cost incurred and of the benefit gained with the change. Panel b: The decision is merely utilitarian. If the benefit is not overcoming the cost, then individuals maintain their current neighborhood (left picture). If, on the contrary, the payoff exceeds the cost, then individuals relinquish their current neighborhood and move to the new one (right picture). The structure of the network then evolves until converging to its Nash equilibrium (if it exists) i.e., to the configuration where no changes of neighborhood are allowed, as no individual has anything to gain in abandoning acquaintances.}
	\label{gametheory}
\end{figure}

Consider the general case which is schematically depicted in Figure \ref{gametheory}, where the $N$ nodes of a network $V$ are rational agents of a game. At each step $m$ of the game, each agent $v \in V$  selects (independently of the choices made by the other agents at the same step) a potential neighborhood ${\cal N}_v (m)$ made of $k_v(m)$ other nodes of $V$. The agent then decides whether it is more profitable to form connections with the nodes in ${\cal N}_v (m)$ or to remain connected with the nodes in ${\cal N}_v (m-1)$. The decision is based on a balance between the payoff and the cost functions associated with the change of neighborhood.

As for the cost function, we assume that node $v$  pays a unitary cost $c>0$ to maintain a connection with each node $u$ belonging to its neighborhood (and that node $u$ cannot refuse the connection paid for by $v$). Moreover, to be as generic as possible, we either assume the unitary cost to be a constant, or to depend on the network size as $c=c(N)$.

\begin{figure}[t!]
	\includegraphics[width=1\linewidth]{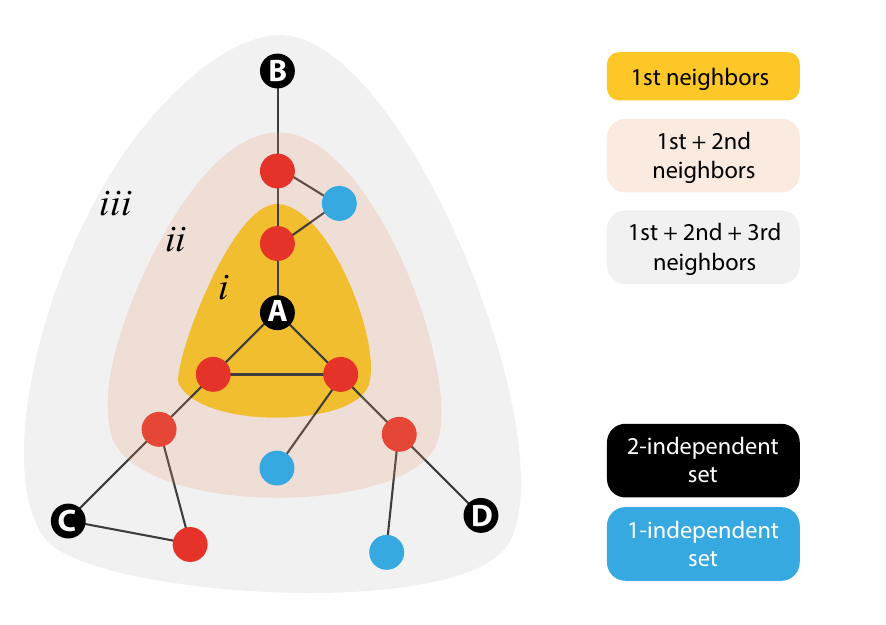}
	\caption{{\bf $l$-independence of nodes.} Sketch of a generic graph, with node A at the center. The first, second and third neighbors of node A are respectively located within the yellow, pink, and gray region. The $l$-independent set of a graph is the set of nodes such that the distance between any two of them is larger than $l$. The black nodes (A, B, C and D) form the 2-independent set of the graph, as all of them are at a distance larger than 2 from each other. The black nodes together with the ones depicted in light blue form, instead, the 1-independent set. Note that the light blue nodes do not participate in the 2-independent set. Finally, the red nodes belong neither to the 1-independent set nor to the 2-independent set.}
	\label{independent}
\end{figure}

As for the benefit function, if agents are rational, it is logical to assume that their goal is to increase their importance within the network. This can be naturally framed in terms of betweenness centrality~\cite{freeman}, which indeed provides a measure of the influence exerted by a node on the information flow within a network. This is defined as follows. First of all, if $v$ and $s$ are two nodes of a connected network, the distance $l(v,s)$ is taken to be the number of edges forming the shortest path between them.
Then, the betweenness centrality (or degree of mediation) $BC(v)$ is taken to be $\sum_{s\ne v\ne t} \frac{\sigma_{st}(v)}{\sigma_{st}}$, where $s,t\in V$ are all possible pairs of different vertices that do not match with $v$, $\sigma_{st}(v)$ is the number of shortest paths between the vertices $s$ and $t$ passing through the vertex $v$, and $\sigma_{st}$ is the total number of shortest paths between the vertices $s$ and $t$.

$BC(v)$ quantifies how relevant the intermediary role played by $v$ in the graph is. However, one immediately realizes that the contribution in $BC(v)$ of the shortest paths in which $v$ is {\it the unique intermediary} between $s$ and $t$ is equal to that of paths in which $v$ is just one of a long chain of intermediaries. To account for such a difference, one may adopt a generic weighted version of the betweenness centrality, $WBC(v)$, which is defined as
\begin{equation}
	WBC(v)=\sum_{s\ne v\ne t}\frac{\sigma_{st}(v)}{\sigma_{st}}\cdot f(l(s,t)),
	\label{benefit}
\end{equation}
where $f$ is a strictly decreasing function of its argument (as longer paths must contribute less).
One can think of Eq.~(\ref{benefit}) as follows: each pair $s, t$ of vertices creates some utility, which is then distributed equally among all shortest paths from $s$ to $t$, and then each intermediary vertex in each path obtains a fraction equal to $\frac{f(l(s,t))}{\sigma_{st}}$.

With these simple rules in mind, the $N$ agents play the game. When the game converges to a Nash equilibrium (a configuration where no agent  has anything to gain by changing its own neighborhood, as all of them have already attained their optimal adjacency), we can demonstrate rigorously that  the obtained structure {\it is endowed with} the six degrees of separation attribute.

\subsection{2-independent sets and the emergence of ultra-small world states}

Before we demonstrate our main results, we first need to introduce the concept of 2-independence of network's nodes.
In traditional graph theory, a 1-independent set (or internally stable set, or anti-clique) $S$ is a set of vertices such that any pair of them isn't connected by a graph's edge.
This is to say that each edge in the graph has at most one endpoint in $S$.
As a consequence, any two vertices of $S$ are at a distance which is strictly larger than one.

One can now generalize the latter definition, and designate as a $l$-independent set $S_l$ the set of network's nodes such that the distance between any pair of its members is larger than $l$~\cite{fink1985}. It follows that nodes belonging to $S_l$ do not necessarily belong to $S_{l+1}$ (see Fig.~\ref{independent} for an illustrative sketch of the comparison between a 2-independent set and a classical 1-independent set).

\begin{figure}[t!]
	\includegraphics[width=1\linewidth]{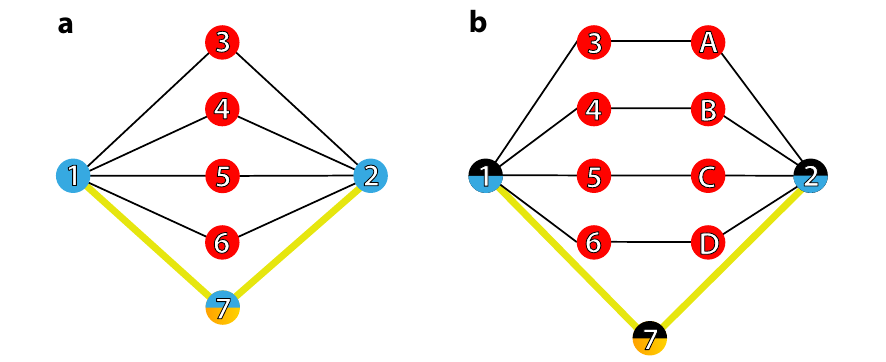}
	\caption{{\bf Independence of nodes and Nash equilibrium.} Panel a:  When only black links are considered, vertices 1,2,7 form a 1-independent set. For consistency with Fig.~\ref{independent}, nodes 1 and 2 are colored in light blue. As vertex 7 forms the yellow edges (7,1) and (7,2) it is removed from the 1-independent set (this change is depicted by coloring the lowest part of the node in yellow), but the two new connections do not remove nodes 1 and 2 from the 1-independent set, since they only contribute to the multiplicity of the shortest paths between 1 and 2.  Panel b: When only black links are considered, vertices 1,2,7 form a 2-independent set. As the yellow connections are formed, vertex 7 reduces the distance between nodes 1 and 2 from at least 3 down to 2. As a consequence, nodes 1 and 2  can only be part of a 1-independent set. For this reason, the upper half of nodes 1 and 2 is depicted in black, indicating that these nodes initially belonged to the 2-independent set, and the lower half in light blue, indicating that by receiving the connections from node 7, they become members of a 1-independent set. Node 7 is half colored in black (as it initially belonged to the 2-independent set), and half in yellow (as the two new connections remove it from all independent sets).}
	\label{sizes}
\end{figure}

Why are 2-independent sets important in our framework?
This can be understood by looking at Fig.~\ref{sizes}.
In panel a, the three vertices 1,2,7 are originally part of a 1-independent set. Now, if vertex 7 forms the yellow edges $(7,1)$ and $(7,2)$, it is removed from the set but it does not change the distance between nodes 1 and 2. It only contributes to the multiplicity of shortest paths between nodes 1 and 2. As the number of alternative shortest paths may be very large in large sized networks, the minimum possible benefit obtained from gluing a 1-independent set (as node 7 would do by forming edges with nodes 1 and 2) may be very small with the growth of the network's size.
From the latter point it follows that the presence of independent sets of large size may be compatible with the Nash equilibrium.

A totally different situation occurs when we consider 2-independent sets, as in panel b of Fig.~\ref{sizes}. Indeed, when forming the yellow connections with nodes 1 and 2, vertex 7 is actually reducing their distance from at least 3 down to 2. Therefore, regardless of which other edge exists in the network involving vertices 1 and 2, vertex 7 receives a minimum benefit equal to $f(2)$. This is equally valid for any other vertex of the 2-independent set which would form edges with all other members of the set: it would receive at least the same benefit from each pair of nodes in the set. Therefore, the minimal benefit obtained from gluing a 2-independence set of size $x$ is ${x - 1\choose 2}f(2)$, which may be rather substantial. For this reason, sizeable 2-independent sets cannot exist in the Nash equilibrium.

The process of gluing large size 2-independence sets is precisely what regulates the spontaneous emergence of the six degrees of separation.
Namely, it can be proved theorematically that such a process determines that
\begin{itemize}
	\item [{\it i)}] at the Nash equilibrium the graph necessarily contains  at least a vertex $v$ whose degree $k$ scales as
	the cube root of the system's size i.e., $k \sim \sqrt[3]{N}$, and
	\item [{\it ii)}] node $v$ is at the center of the network and displays the remarkable property
	of being at a distance of no more than 3 from any other node of the graph.
\end{itemize}
The latter implies that the shortest path between any pair of nodes $i,j$ in the graph will be smaller or equal than 6, as there will be maximum 3 edges forming the shortest path from $i$ to $v$ and maximum three edges also to form the shortest path from $v$ to $j$. Therefore, the diameter $D$ of the network will be exactly 6.

The proofs of the Theorems and Lemmas involved are available in the Supplementary Information Appendix (SI).

\subsection{An illustrative case}

For the sake of a better illustration , let us now
focus on the case described as follows.
\begin{itemize}
	\item [{\it i)}] The agents start the game when they are already connected by means of a pristine graph where, in addition,
	there exists at least one node with sufficiently high degree.
	\item [{\it ii)}] Each agent $v$ adopts as benefit function
	\begin{equation}
		WBC(v)=\sum_{s\ne v\ne t}\frac{\sigma_{st}(v)}{\sigma_{st}}\cdot \frac{1}{l(s,t)^\alpha},	
		\label{benefit2}
	\end{equation}
	with $\alpha$ being a strictly positive parameter. Comparing with Eq.~(\ref{benefit}), this means that the weighting factor is $f(l(s,t))=\frac{1}{l(s,t)^\alpha}$,
	and that Eq.~(\ref{benefit2}) coincides, for $\alpha=1$, with the classical weighted betweenness centrality \cite{freeman}.
	\item [{\it iii)}] Agents sequentially add new connections to their
	neighborhood if and only if there is a positive balance between the extra-utility brought by the new connections and the extra-cost.
\end{itemize}

In practice, at each step $m$ of the game, the potential neighborhood ${\cal N}_v (m)$ of each agent
$v \in V$  is equal to ${\cal N}_v (m-1)$ plus $p$ other nodes.
The $p$ new edges are then added only if $\Delta WBC(v) \geq pc$ i.e., only if the extra weighted betweenness centrality
is larger or equal to the extra cost $p c$.

When no agent is able to incorporate any further edge, the network is said to have reached its asymptotic equilibrium.
It should be remarked that such a final state cannot formally be associated to a Nash equilibrium, because the option of removing existing links
is not contemplated in the game, and therefore there is no certainty that agents, in their asymptotic states, are in their optimal adjacency configuration.
In this respect, it is worth highlighting that another mechanism (beyond that of link addition and deletion) that one can consider at the basis of
the emergence of the six degrees of separation is that of link rewiring, which would actually imply the invariance of the network density during its evolution towards
the asymptotic equilibrium. We plan to report on the effects of this latter mechanism elsewhere.

The following Theorem can be proved:
\begin{itemize}
	\item [{\it i)}] if $v$ is a node of the pristine graph with $k$ original connections,  and
	\item [{\it ii)}] if $H \in \left\{3,4,5,...\right\}$ is some integer number strictly larger than 2, and
	\item [{\it iii)}]  if, for the considered values of $c$ and $\alpha$, the inequality
	\begin{equation}
		\label{3}
		\left(\frac{1}{2^\alpha}-\frac{1}{(H+2)^\alpha}\right)k \geq c,
	\end{equation}
	is satisfied,
\end{itemize}
then, in the equilibrium state of the network, the node $v$ is linked to all other nodes of the graph by no more than $H$ links, implying that the diameter of the equilibrium network does not exceed $2 H$.

In practice, the theorem guarantees  that the asymptotic state of a network evolving from an initial condition that satisfies condition~(\ref{3}) is an ultra-small world state (and, for $H=3$, also the emergence of the 6 degrees of separation property).

\begin{figure}
	\includegraphics[width=1\linewidth]{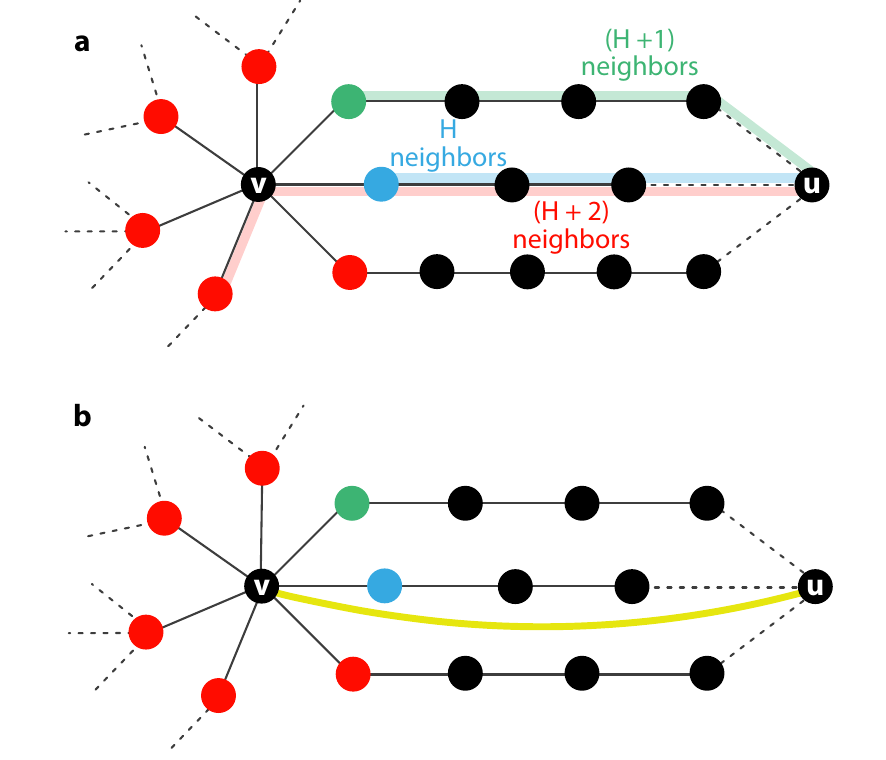}
	\caption{{\bf The emergence of the ultra-small world state.} Panel a: Sketch of a hypothetical network where nodes $v$ and $u$ are separated by a distance $H+1$. The  neighbors of $v$ are then at either $H$ (the light blue node), or $H+1$ (the green node $s$), or $H+2$ (the red nodes) edges from $u$. For a better visualization, paths of different lengths are marked with the corresponding colors.  Panel b: A direct (yellow) link is added between $v$ and $u$. Our study demonstrates rigorously (see Theorem 3 of the SI) that the network configuration of panel a is incompatible with an equilibrium state. Notice that, at variance with the case depicted in Fig.~\ref{independent},
		here distances between nodes depend on the value of the parameter $H$.}
	\label{theorem}
\end{figure}

The proof of the Theorem (see SI for details) is given by contradiction i.e., by supposing that there is a node $u$ in the final state of the network whose distance from $v$ is at least $H+1$ i.e., $l(u,v)\geq H+1$. To better illustrate the situation, we depicted in panel a of Fig.~\ref{theorem} the case in which nodes $v$ and $u$ are separated by a distance $H+1$. In that circumstance, the nodes directly connected to $v$ (the neighbors of $v$) may be found at either $H$ (the light blue node), or $H+1$ (the green node), or $H+2$ (the red nodes) edges from $u$.
Looking at the figure, it is easy to understand that all network's shortest paths which end in $u$ and start in either the green or the light blue node {\it cannot} pass through $v$. Therefore, the only contribution  to the benefit function of $v$ from shortest paths ending in $u$ is coming from those paths which start in the red nodes, the neighbors of $v$ that are at distance  $H+2$ from $u$.

When one, instead, includes a direct link between $v$ and $u$ [the yellow link in panel b of Fig.~\ref{theorem}], then the shortest path between any neighbor of $v$ (denoted by $w$) and $u$ becomes $w-v-u$, since $H\geq 3$. Calculating then the value of $\Delta WBC(v)$ corresponding to the addition of such a link, and recalling that the equilibrium requires $\Delta WBC(v)$ to be smaller than the cost $c$, one easily get to an expression which is in explicit contradiction with condition~(\ref{3}) (see the SI for full details).

\subsection{A realistic case}

We shall remark that our approach is valid independently on the specific degree distribution properties of the pristine graph.
However, the maximum degree of a node in a scale-free network generated by the preferential attachment method~\cite{barabara}
is known to scale as $\sqrt{N}$~\cite{BR,FFF} and this implies that, for these networks, condition~(\ref{3}) is (from a given size on) always verified for any value of fixed cost $c$ and any value of $\alpha$, thus making them very good candidates for initializing the formation of ultra-small world structures.

Therefore, to illustrate power and generality of the above Theorem, we performed a massive numerical trial by initializing our game on networks of $N$ nodes generated with the Barab\'asi-Albert (BA) algorithm~\cite{barabara}, for $\alpha=1$ (i.e., adopting as benefit the weighted betweenness centrality), $H=3$ and $c=0.15\sqrt{N}$ (to ensure a coherent scaling of the cost with that of the maximum degree in the network).
With these stipulations, condition~(\ref{3}) becomes $0.3 k\geq 0.15 \sqrt{N}$. As $k\approx 2\sqrt{N}$~\cite{BR,FFF}, this means that condition~(\ref{3}) is verified at each value of $N$, and one then expects that the diameter at equilibrium would not exceed $6$.

It is important to remark here that estimating the benefit function~(\ref{benefit2}) requires the retrieval of the global structure of network's pathways at each step of the game. However, such information is in general not available to the agents of real social networks.
Indeed, computing Eq.~(\ref{benefit2}) becomes prohibitively costly as the size of the network increases, requiring (with the fastest existing algorithms) ${\cal O} (NL)$ operations (being $L$ the total number of links in the network)~\cite{newma,brand}.

For this reason, it is much more realistic and much less computationally demanding to assume that agents only use local information. We then consider a scenario wherein at each step $m$ of the game, a (large degree) node $v$ is chosen. $v$ incorporates an edge with another node $u$ if
\begin{enumerate}
	\item [a)] $0.3 k \geq c$, where $k$ is the degree of $v$,
	\item [b)] the distance between $u$ and $v$ is larger than $3$.
\end{enumerate}

In this way, it is only required to check that the subgraph formed by $v$ and its first and second neighbors has zero overlap with the subgraph formed by $u$ and its first neighbors, and the method is not hurting for the knowledge of the overall shortest paths' structure.
At the same time, the adoption of local information makes our study's main claims even stronger, because it proves that a global network property (the network diameter)
may emerge as a result of a game in which agents only share local information, which is what actually happens in almost all real circumstances.

\begin{figure}[t!]
	\includegraphics[width=0.99\linewidth]{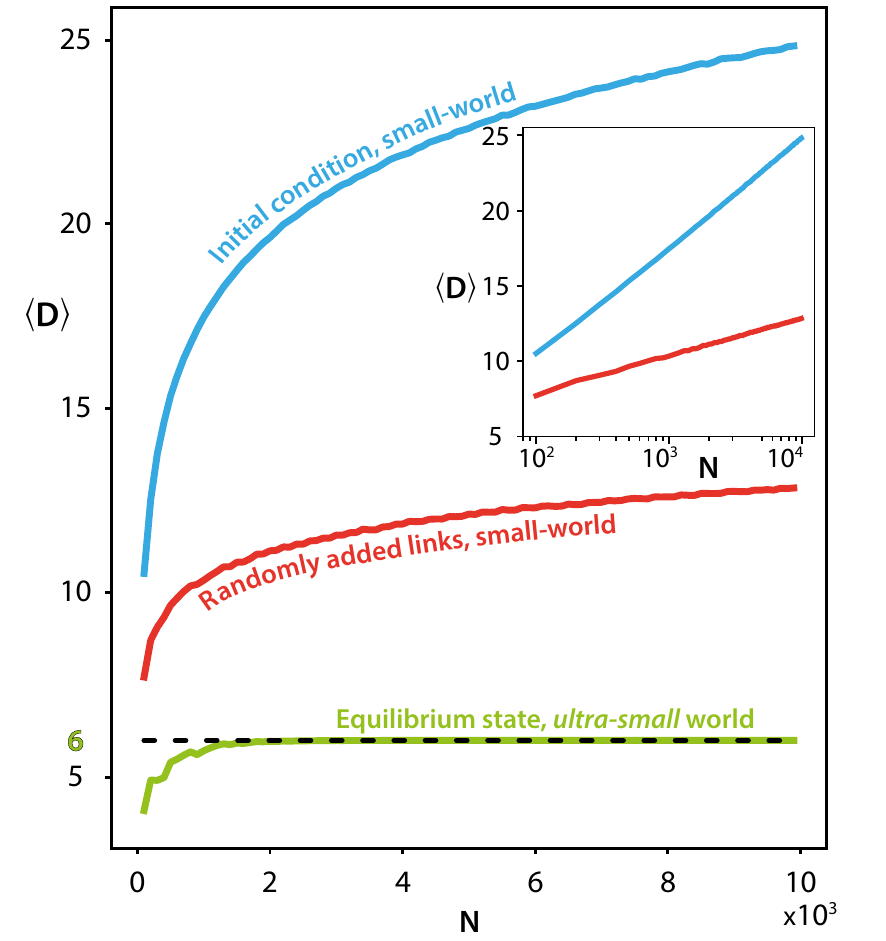}
	\caption{{\bf The emergence of the six degrees of separation.} Ensemble average $\langle D \rangle$ vs. $N$ for different sets of networks. Light blue line: BA scale-free networks that are used as initial conditions for the evolution of the game. Green line: Networks generated at the equilibrium state of the game. Red line: Networks constructed by randomly adding to the initial condition of each game the same number of links needed to reach the game equilibrium.
		A horizontal black dashed line is positioned at $D=6$ to indicate that the network's structure obtained at the equilibrium features the ultra-small world property,  with the concurrent emergence of the six degrees of separation.
		Inset: log-lin plot of $\langle D \rangle$ vs. $N$. The logarithmic scaling of the light blue and red lines is clearly visible.}
	\label{results}
\end{figure}

Note that, if an edge connecting $u$ and $v$ is added, the above conditions implies that $\Delta WBC(v) \geq c$. Indeed,  if the node $u$ satisfies condition $b)$, it can  easily be shown (using the same arguments that the reader finds in the SI for the proof of the Theorem) that the maximum contribution to $v$ of the shortest paths between $u$ and a neighbor of $v$ is $1/5$. Adding the new edge, such a contribution raises to $1/2$, and this means that
\begin{equation*}
	\Delta WBC(v)\geq 0.3k,
\end{equation*}
where $k$ is the number of connections of $v$. Therefore, if condition $a)$ holds, then condition $\Delta WBC(v) \geq c$ is also satisfied. This implies that our local method is actually more restrictive when incorporating edges, and yet sufficient to give evidence of the predictions of the Theorem for nodes satisfying condition~(\ref{3}).

The results of our simulation trial are presented in Fig.~\ref{results}. At each value of the network size $N$, $10,000$ different realizations of a BA scale-free network are generated. The ensemble average $\langle D \rangle$ of the value of the network diameter is plotted as a light blue line in the figure, showing a small-world behavior (a logarithmic scaling with $N$, well visible in the log-lin plot of the inset).

Each of the generated networks is then taken as initial condition for the evolution of our game, following the conditions a) and b) described above, until reaching the final, equilibrium state. $\langle D \rangle$ for the reached equilibria is reported as a green line in the figure, and it is clearly seen that an ultra-small world state emerges with $\langle D \rangle=6$ (a value marked by an horizontal dashed line).

A legitimate objection is that adding links to a graph (and therefore increasing the graph's density) always results in decreasing the network's diameter, and therefore a proper comparison has to be offered to assess the relevance of the obtained results.
For this purpose, in all trials we took diligent note of the total number of links added before reaching the equilibrium. Then, we took back the initial condition of the specific trial, and added exactly the same number of links, but this time in a fully random way i.e.,  without caring about the fulfillment of the game conditions a) and b).
The obtained values of $\langle D \rangle$  are reported as a red line in the figure. As expected, the red line is always located below the light blue one, but the remarkable result is that the new network ensemble maintains exactly the same logarithmic scaling with $N$ (once again well visible in the inset), which is destined to depart more and more from the constant value characterizing ultra-small world states and emerging at the equilibrium of our game.

\begin{figure}[t!]
	\includegraphics[width=0.99\linewidth]{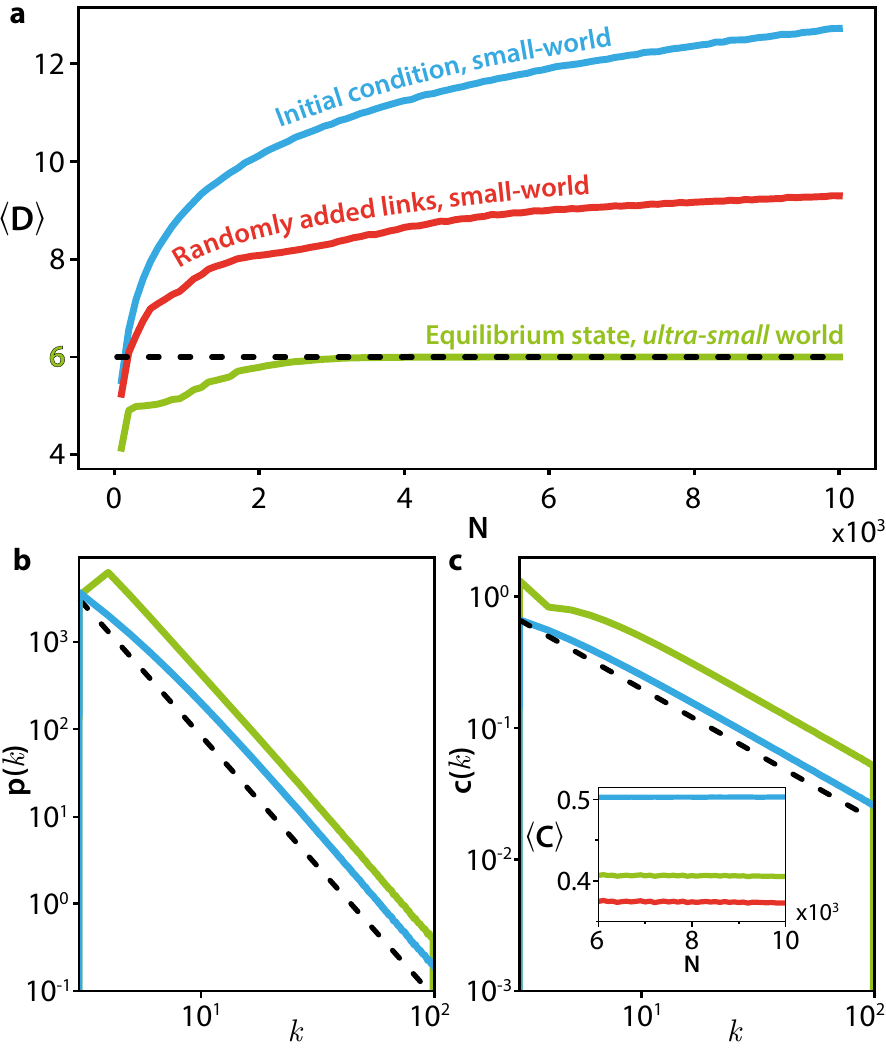}
	\caption{{\bf Scale-free distribution and hierarchical clustering.} a) Ensemble average $\langle D \rangle$ vs. $N$ for the three considered ensembles of networks. Light blue line: networks generated by the procedure of Ref.  \cite{boccalettibravo} and  that are used as initial conditions for the evolution of the game. Green line: The equilibrium network states of the game. Red line: Networks constructed by randomly adding to the initial condition of each game the same number of links needed to reach the game equilibrium. A horizontal black dashed line is positioned at $D=6$. b) The degree distribution $p(k)$ vs. $k$ for the set of initial conditions (light blue line) and the set of reached equilibria (green line). $N=10,000$.  For visibility, the green line plotting $p(k)$ at the equilibria has been slightly vertically shifted. The black dashed line reports the scaling $p(k) \sim k^{-3}$. c) The hierarchical clustering coefficient $c(k)$ (see text for definition) vs. $k$ for the set of initial conditions (light blue line) and the set of reached equilibria (green line). $N=10,000$. For visibility, the green line plotting $c(k)$ at the equilibria has been slightly vertically shifted. The black dashed line reports the scaling $c(k) \sim k^{-1}$. The small inset reports the global clustering coefficients $\langle C\rangle$  vs. $N$ for the three considered ensembles.
	\label{clustering}
}
\end{figure}

Finally, we move to show that the mechanism proposed by us and leading to the emergence of the six degrees of separation is, in fact, perfectly compatible with all major structural properties that are observed in real social networks, and in particular with scale-freeness in the degree distribution and with the presence of prominent and hierarchical clustering features. The former attribute indicates that the distribution of the nodes' degrees scale as $p(k) \sim k^{-\gamma}$ (with $2 < \gamma \leq 3$ in real social networks), the latter implies that the clustering coefficient $c(k)$ of a connectivity class $k$ (the average clustering coefficient of all nodes with a given degree $k$) does depend on $k$ as $c(k) \sim k^{-\omega}$~\cite{ravasz2002}.

To that purpose, we repeat the same extensive simulations which led us to obtain the results reported in Fig.~\ref{results}, but this time we adopted as initial conditions for each trial networks that are originated by means of the procedure described in Ref.~\cite{boccalettibravo}, which indeed provides graphs endowed with degree distributions $p(k) \sim k^{-3}$, with a very high clustering value ($c \sim 0.5$ for an average degree of $\langle k \rangle=6$), and with a hierarchical structure of the clustering described by $c(k) \sim k^{-1}$.

Once again, for each value of $N$, an ensemble of $10,000$ different networks are synthesized by the technique of Ref.~\cite{boccalettibravo}, and each of the generated networks is taken as initial condition for the evolution of the game, until reaching the equilibrium state. In each trial, moreover, note is taken of the total number of links added before reaching the equilibrium, and a network is constructed, for comparison, by randomly adding exactly the same number of links to the used initial condition.

The results are reported in Fig.~\ref{clustering}. Precisely, panel a of Fig.~\ref{clustering} clearly shows that the scenario obtained is identical
to that of Fig.~\ref{results}: the values of $\langle D \rangle$ averaged over the ensemble of the initial conditions (light blue line) and over the constructed set of networks with randomly added links (red line) are both scaling logarithmically with $N$, while the reached equilibria (green line) are ultra-small world states with $\langle D \rangle=6$ (marked by an horizontal dashed line).

Panels b and c of Fig.~\ref{clustering} compare, instead, the structures of the initial conditions and that of the reached equilibria, for $N=10,000$, and one immediately sees a very remarkable fact: {\it all other} structural properties imprinted in the initial conditions are conserved in the final state.  Precisely, panel b (panel c) of Fig.~\ref{clustering} reports the degree distribution $p(k)$ (the clustering coefficient $c(k)$) for the light blue case corresponding to the used initial conditions and for the green case corresponding to the reached equilibria, and one immediately sees that the scaling $p(k) \sim k^{-3}$ ($c(k) \sim k^{-1}$), highlighted by a black dashed line, is fully preserved within the range $10^0-10^2$ of the degree i.e., across two orders of magnitude, and with minimal differences occurring only at larger degrees due to the addition of the new links that create a few new hubs at the equilibrium. For visibility, $p(k)$ ($c(k)$) of the equilibria has even been multiplied by 2, in order to shift the line in the panel, because otherwise there would be an almost complete overlap between the light blue and the green curves.

In the small inset of panel b of Fig.~\ref{clustering}, the values of the global clustering coefficients are reported vs. $N$ for the three ensembles. One sees that the addition of links in the process of relaxation to equilibrium leads to a slight decrease of $\langle C\rangle$ (from $\langle C\rangle \sim 0.5$ to $\langle C\rangle \sim 0.42$) which, however is maintained to a level pointing to the presence of very prominent and important clustering features. However, the most remarkable trait here is that the value of $\langle C\rangle$ at equilibria is {\it larger} than that pertinent to the ensemble of networks constructed by randomly adding to the initial conditions the same number of links needed to reach the equilibria.

\section{Discussion and outlook}

The compensation of costs and benefits is certainly a natural interaction strategy through which rational agents determine their connections~\cite{eguiluz2005cooperation, perc_bs10, jackson2010social, christakis2013social, nishi2015inequality, alvarez2021evolutionary}, and therefore our study contributes to the understanding of why the six degrees of separation is such a ubiquitous property across vastly different social networks.
It is, moreover, reasonable to assume that a similar evolutionary principle may also apply to the design of man-made or technological networks~\cite{holme2012temporal}: take, for instance air or sea transportation networks~\cite{guimera2005worldwide, barbosa2018human, lei2022forecasting}, in which airports/ports may increase their volume of trades and/or tourism industry by `being in between' the main routes of interchanges of passengers and goods, and in doing so they are keen to incur in the relative costs of maintaining (or even enlarging) the number of  local connections.

On the other hand, the units of biological networks are in general not rational agents, and it is not straightforward to argue that benefits in terms of betweenness centrality shaped, for instance, the structure of metabolic, genetic or brain networks along their million year long evolutionary path~\cite{jeong2000large, bassett2017network, zwir2022evolution, tripp2022evolutionary}. However, one cannot rule out that other compensation mechanisms could have played a pivotal role in this case too, with different benefit functions (e.g. resilience to random perturbations or failures~\cite{albert2000error, cohen2000resilience}, or local or global efficiency~\cite{morone2015influence}) recouping for the cost to form or maintain a specific adjacency structure. In neural structures, for instance, it is well known that the functional gains associated with link formation must offset the associated structural costs~\cite{cajal,bullmore,sterling}. Note that, for neural structures, while this principle holds in general at evolutionary and developmental time scales, it may also take place at much shorter scales, comparable to those of social network dynamics.

Finally, our study also sheds light on the so-called {\it strength of weak ties} phenomenon. This concept was introduced by Mark Granovetter who showed that the most common way of finding a new job is through personal contacts with distant acquaintances, and not via close friends, as one would instead have expected~\cite{granovetter1973strength,granovetter1974}. Distant acquaintances represent links connecting different groups of people, and therefore provide each individual with a unique way to receive useful information about distant groups.

Formally speaking, weak ties are links connecting nodes that were originally located at rather large distances and they are therefore called {\it bridges} or {\it local bridges} (see the discussion and references in Chapter 3 of Ref.~\cite{easley2010networks}). Their importance for social interaction and communication is strongly supported by a wide range of studies~\cite{burt1992structural,burt2004structural}.

The formation of links connecting nodes from 2-independent sets as the key to the emergence of the six degrees of separation describes exactly the case of a local bridge formation, i.e. a {\it weak tie} in Granovetter's sense. Therefore, our model can also be viewed as the game-theoretical foundation for the {\it strength of weak ties} phenomenon.

Authors would like to thank Gonzalo Contreras-Aso and Jorge Tredicce for many inspiring discussions. R. Criado and M. Romance acknowledge funding from projects PGC2018-101625-B-I00 (Spanish Ministry, AEI/FEDER, UE) and M1993 (URJC Grant). M.P. was supported by the Slovenian Research Agency
(Grant Nos. P1-0403 and J1-2457). The usage of the resources, technical expertise, and assistance provided by the supercomputer facility CRESCO of ENEA in Portici (Italy) is also acknowledged.


\newpage

\begin{titlepage}
	\begin{center}
		\Large\textbf{Supplementary Information: \\ \textit{Why are there six degrees of separation in a social network?}}\\ \vspace{0.5cm}
		\large{I. Samoylenko$^*$, D. Aleja$^*$, E. Primo, K. Alfaro-Bittner, E. Vasilyeva, K. Kovalenko, D. Musatov, A. M. Raigorodskii, R. Criado, M. Romance, D. Papo, M. Perc, B. Barzel, S. Boccaletti.}
	\end{center}
\end{titlepage}

\onecolumngrid

	
In this Supplementary Material (SM), the reader finds all details of the theorematic proofs which are referred to in the main text.
The SM is divided in two main sections. The first Section contains some definition and preliminaries that are of use in all
Theorem and Lemmas and describes the results of the game theoretical approach presented in the first part of our Manuscript.
The second Section contains, instead, the details of the illustrative case (presented in the second part of our Manuscript)
in which nodes can only add links to their neighborhoods.

\section{The game theoretical approach}

In our model, the $N$ nodes of a network $V$ are agents of a game.
At each step $m$ of the game, each node $v \in V$ selects (independently on the choices that may be made by the other agents in the same step)
a potential neighborhood ${\cal N}_v (m)$ made of $k_v(m)$ other nodes of $V$, and decides whether it is more profitable to form connections with all the nodes in ${\cal N}_v (m)$ or to remain connected with the nodes in ${\cal N}_v (m-1)$. The decision is based on compensation between the costs incurred in the change and the payoff (or benefit).

To be more specific,  at the step $m$ of the game,  each agent $v$ (separately and independently) compares the configuration defined by the adjacency matrix $A_{m-1}$ (reflecting the state of the network after all agents have made their choice of neighborhood at the step $m-1$)
with that of the adjacency matrix $A_{m,u}$ which is obtained from $A_{m-1}$ by
eliminating all the $k_v(m-1)$ edges between $v$ and the members of ${\cal N}_v (m-1)$ and adding instead all the $k_v(m)$ edges between $v$ and the members of ${\cal N}_v (m)$.
If the benefit of the new configuration $A_{m,u}$ overcomes the costs of forming such $k_v(m)$ edges, then the agent $v$ adopts the neighborhood ${\cal N}_v (m)$, otherwise it remains linked with the members of ${\cal N}_v (m-1)$.
The step $m$ of the game is concluded when all the $N$ agents have made their decision and a new global network arrangement is produced, reflected by the adjacency matrix $A_{m}$, which is then used (by all agents) at the step $m+1$. \\

\subsection{Definitions and preliminaries}

\textbf{Definition 1} The cost of forming a connection  $(u,v)$ is taken to be equal to some $c (N) > 0$, which can  be either a constant or a generic function of the network size. \\

\textbf{Definition 2}. The distance between two vertices is the number of edges forming the shortest path between them.\\

\textbf{Definition 3}. The benefit function is taken to be $WBC(v)=\sum_{s\ne v\ne t}\frac{\sigma_{st}(v)}{\sigma_{st}}\cdot f(l(s,t))$ [as in Eq. (1) of the main text], where $s,t\in V$ are all possible pairs of vertices other than $v$, $\sigma_{st}(v)$ is the number of shortest paths between the vertices $s$ and $t$ passing through the vertex $v$, $\sigma_{st}$ is the total number of shortest paths between vertices $s$ and $t$, $f(x)$ is an arbitrary (but strictly decreasing) function of the argument $x$, and $l(s,t)$ is the length of the shortest path between vertices $s$ and $t$. \\

\textbf{Definition 4}. A network satisfies the \textit{ultra-small world} property if its diameter (the maximal distance between any pair of network's nodes) is bounded by a given value which is independent on the network's size. \\

\textbf{Definition 5}. A $2$-independent set $S$ is a set of network's nodes such that the distance between any pair of its members is larger than $2$. As a consequence, each pair of nodes $u,v \in S$ isn't connected directly by a network edge, nor a vertex $w \in V$ exists having simultaneously connections with $u$ and $v$. \\

It is now necessary to make a couple of preliminary observations, that are of use in all demonstrations which are part of this SM.

A first observation is concerned with the fact that, if a node $v$ selects a potential new neighborhood where all previous connections are maintained and $p$ new neighbors are added and if such a new configuration is not accepted, it actually implies
\begin{equation}
	\label{diferencia}
	\Delta WBC(v)< p c,
\end{equation}
where $\Delta WBC(v)$ is the difference between $WBC'(v)$ (calculated with incorporating the $p$ new edges) and $WBC(v)$.
Indeed, as the new strategy is rejected, it follows that
\begin{equation*}
	WBC'(v)<(q+p)c,
\end{equation*}
where $q$ was the number of neighbors of $v$ before incorporating the $p$ new edges. Moreover, as the $q$ previous neighbors were instead accepted in the previous step of the game, one has that $WBC(v)\geq q c$. Therefore, condition~\eqref{diferencia} comes from the fact that
\begin{equation*}
	WBC'(v)<(q+p)c\leq  WBC(v)+pc.
\end{equation*}

A second, important, observation comes from the content of the following Lemma 0, which states that the contribution to the benefit function of $v$ given by the shortest path between any two nodes (say $s$ and $t$) never decreases when $v$ acquires a new link.
Fig. 1A illustrates the case in which node $v$ accrues an utility due to being intermediary in one of the shortest paths connecting nodes $s$ and $t$.
If a link is now added between $v$ and another node $u$, the benefit function for $v$ changes as
\begin{equation}
	\label{SM2}
	WBC'(v)=\sum_{s\neq v\neq t} \frac{\sigma'_{st}(v)}{\sigma'_{st}}\cdot  f(l'(s,t)).
\end{equation}
where now all quantities are denoted by $'$. Notice that the number of pairs $(s,t)$ in Eq.~\eqref{SM2} may change from the ones considered in $WBC(v)$ because of the added link. \\

\textbf{Lemma 0}.
If a new link is added between two nodes $u$ and $v$ then

\begin{equation}
	\label{SM3}
	\frac{\sigma_{s,t}(v)}{\sigma_{s,t}}\cdot f(l(s,t))\leq  \frac{\sigma'_{s,t}(v)}{\sigma'_{s,t}} \cdot f(l'(s,t)),
\end{equation}
where $s$ and $t$ are any two nodes of $V\setminus \left\{v\right\}$. \\

\begin{figure}[t]
	\centering
	\includegraphics[scale=1]{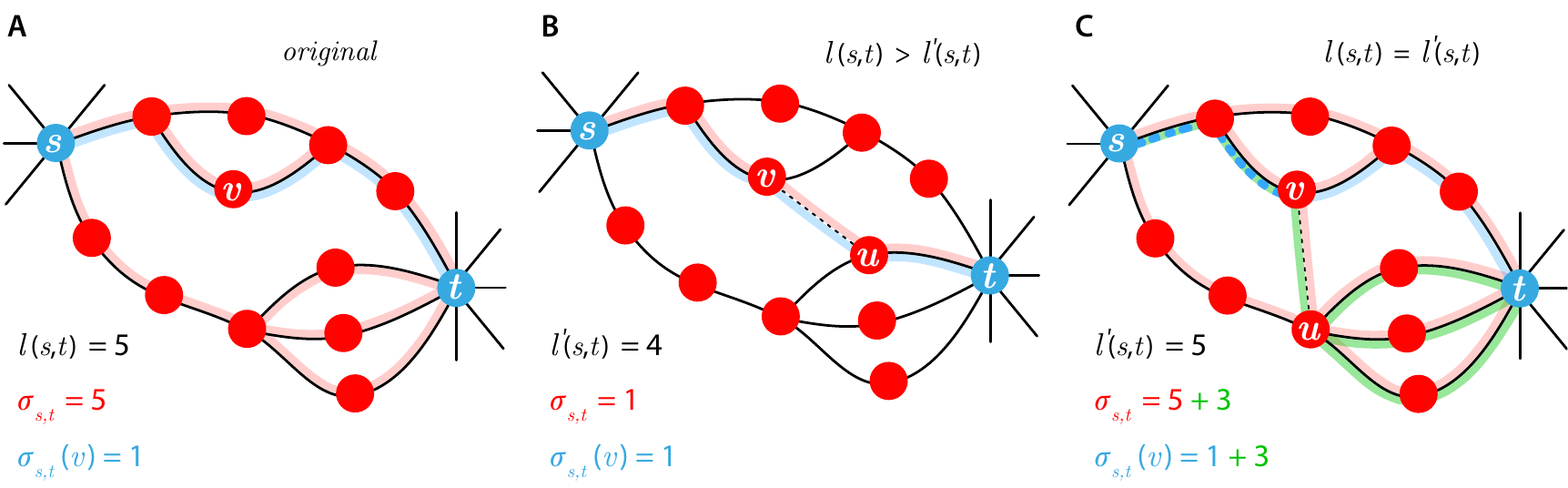}
	\caption{Schematic illustration of the two possible cases examined in Lemma 0. In all panels red and light blue curves are used to mark, respectively,  the shortest paths between $s$ and $t$ (whose number is $\sigma_{s,t}$) and the shortest paths between nodes $s$ and $t$ which pass through node $v$ (whose number is $\sigma_{s,t} (v)$). {Panel A)} shows an hypothetical graph $V$ where nodes $s$ and $t$ (colored in light blue) are separated by a distance $l(s,t) = 5$, $\sigma_{s,t} = 5$, and $\sigma_{s,t}(v) = 1$. The next two panels illustrate the two possible cases originated by adding a link between nodes $v$ and $u$. {Panel B)}  Case $l(s,t) > l'(s,t)$. In this situation, both the distance between $s$ and $t$ and the number of shortest paths decrease. In particular, the unique (new) shortest path between $s$ and $t$ is the one that passes through node $v$. {Panel C)} Case $l(s,t) = l'(s,t)$. In this case three new shortest paths appear (colored in green). Note that all the shortest paths colored in red are equal to the ones presented in panel A.}
	\label{Fig1}
\end{figure}

\textbf{Proof of Lemma 0.}
When adding such a link, the distance between $s$ and $t$ can either decrease or remain the same. Then, these two cases have to be separately examined. \\
\begin{itemize}
	\item Case $l(s,t)>l'(s,t)$ (illustrated in Fig. 1B). In this case, when a link is added between $u$ and $v$, one obtains that
	\begin{equation*}
		\frac{\sigma_{s,t}(v)}{\sigma_{s,t}}\cdot f(l(s,t))\leq f(l'(s,t)),
	\end{equation*}
	because $\sigma_{s,t}(v)\leq\sigma_{s,t}$, \ $l(s,t)>l'(s,t)$, and $f$ is a strictly decreasing function of its argument.  Therefore, as all the shortest paths between $s$ and $t$ pass through $v$ (see again Fig.1B), one has that $\sigma'_{s,t}(v)=\sigma'_{s,t}$, and condition \eqref{SM3} is verified.
	
	\item Case $l(s,t)=l'(s,t)$ (illustrated in Fig. 1C). In this situation, when a link between $u$ and $v$ is added, $x$ new shortest paths between $s$ and $t$ may arise, which however have to pass all through $v$ (as it is shown in Fig.~\ref{Fig1}C). Therefore, one has
	\begin{equation}
		\label{SM4}
		\sigma'_{s,t}=\sigma_{s,t}+x \quad \hbox{and} \quad \sigma'_{s,t}(v)=\sigma_{s,t}(v)+x.
	\end{equation}
	Thus, as $\sigma_{s,t}(v)\leq\sigma_{s,t}$, one obtains that
	\begin{equation*}
		\sigma_{s,t}(v)\left(\sigma_{s,t}+x\right)\leq  \sigma_{s,t}\left(\sigma_{s,t}(v)+x\right),
	\end{equation*}
	and eventually
	\begin{equation*}
		\frac{\sigma_{s,t}(v)}{\sigma_{s,t}}\leq \frac{\sigma_{s,t}(v)+x}{\sigma_{s,t}+x}.
	\end{equation*}
	
	Therefore, considering both $l(s,t)=l'(s,t)$ and Eq.~\eqref{SM4}, it comes out that condition~\eqref{SM3} is satisfied.
\end{itemize}

\subsection{Main results}

Here, we describe the properties of the model's Nash equilibria (when they exist), i.e. of those settings where it is unprofitable for any agent to unilaterally deviate from its current strategy.

First, we state a bounding condition for the cost ensuring that, when they exist, such Nash equilibria are not empty graphs, and therefore they contain at least a connected component. \\

\textbf{Theorem 0.} If $c(N) > \frac{f(2)(N-2)}{2}$, then the empty graph is a Nash equilibrium configuration. Otherwise, an empty graph will never be obtained at the Nash equilibrium.\\

\textbf{Proof.} Consider a vertex $v$ of an empty graph $V$. From each pair of other vertices $s,t\in V, s\ne v\ne t$, the contribution to $WBC(v)$ is no more than $f(2)$ (which would correspond to the case in which both nodes $s$ and $t$ are directly linked to $v$). The net benefit for $v$ of making $x$ connections will be then equal to ${x\choose 2}\cdot f(2) - xc$. Obviously, the maximum is reached at $x = N-1$. Then, if a vertex $v$ in an empty graph changes its strategy and forms connections with the other $N-1$ nodes, it receives a net utility equal to ${N - 1\choose 2}\cdot f(2) - (N - 1)c$. Such a latter quantity will be non negative for $\frac{(N - 1)(N - 2)f(2)}{2} \geq (N - 1)c \Rightarrow c \leq \frac{f(2)(N-2)}{2}$, meaning that a vertex can exist which would benefit from deviating from the strategy of not drawing any edges. If instead $c > \frac{f(2)(N-2)}{2}$, then the maximum possible gain (after deducing costs) will be negative, and this implies that the empty graph will be a Nash equilibrium.\\

\subsubsection{Condition for the existence of a vertex of a large degree}
The next step is to prove that, in a non empty Nash equilibrium, there exists always a vertex of large degree. \\

\textbf{Lemma 1}. Let the maximum degree of a vertex in a connected graph $V$ be $k$. Then, the maximum 2-independent set will be of size at least $\frac{|V|}{k^2+1}$.\\

\textbf{Proof.} Let's construct iteratively a 2-independent set $S$ of the required size.

For this  purpose, we start by considering a generic node $v_1\in V$. Then, we consider the set $t_1$ including $v_1$, all its neighbors (vertices of $V$ which are at distance 1 from $v_1$) and all its neighbors of neighbors (vertices of $V$ at  distance 2 from $v_1$). Since the degree of a vertex in $V$ does not exceed $k$, so does the maximum number of neighbors of $v_1$. Consider now the vertex $u$ which is neighbor of $v_1$. Since its degree does not exceed $k$ and it has a connection $(v,u)$, then the maximum number of vertices that are at distance 2 from $v_1$ and are furthermore connected to $u$ is $k-1$. It follows that the number of vertices belonging to $t_1$ is bounded from above as $k(k-1)+k+1=k^2+1$ (a situation in which $v_1$ has $k$ neighbors and each neighbor of $v_1$ provides a unique set of vertices located at a distance of 2 from it).

The vertex $v_1$ is added to the set $S$ and the subgraph $V_1 = V \setminus t_1$ is considered. $V_1$ is the graph obtained from $V$ by removing all members of $t_1$, and therefore it contains vertices that are all at a distance of at least 3 from $v_1$ (in other words, any vertex $v_2\in V_1$ will be 2-independent with $v_1$). The procedure can be repeated iteratively until there are no vertices left in the resulting subgraph: at each iteration of the procedure, a new vertex can be added to $S$, and no more than $k^2 +1$ vertices are removed from $V$.
As a consequence, there are at least $\frac{|V|}{k^2 + 1}$ iterations of the procedure, which means $|S|\geq\frac{|V|}{k^2 +1}$ ({\it quod erat demonstrandum!}) \\

\textbf{Lemma 2.} Let a 2-independent set of size $x$ be present in the network. Then, if one vertex of such a set "glues the set" (i.e., it forms connections with all other members of the set), it receives an additional utility equal to at least ${x - 1 \choose 2}\cdot (f(2) - f(6))$\\

\textbf{Proof.} Consider a 2-independent set $S$ and a vertex $v\in S$, and let us estimate the minimum gain that such a vertex will get by forming edges to all other vertices of $S$. Notice that, before forming connections in the set, the maximum contribution to $WBC(v)$ received from a pair of other vertices $s, t \in S$ is $f(6)$. This is because both distances $l(v,s)$ and $l(v,t)$  are at least 3, and therefore either the shortest path between $s$ and $t$ is of length at least 6, or it does not pass through $v$ (and, in this latter case, the contribution to $WBC(v)$ is 0). This implies that even in the case in which all shortest paths between $s$ and $t$ are passing through $v$, the maximum possible contribution to $WBC(v)$ from these two vertices is $f(6)$. Now, let the vertex $v$ form the edges $(v,s)$ and $(v,t)$. Then, there exists a unique path from $s$ to $t$ of length 2 passing through $v$ (see the explanatory Figure 2B in the main text), and thus such a pair of vertices will contribute $f(2)$ to $WBC(v)$. On its turn, this implies that the minimum increase of $WBC(v)$ from gluing any pair of vertices $s,t \in S$ is equal to $f(2) - f(6)$, and then (summing up over all possible pairs of vertices $s, t \in S$) the Lemma is proved.\\

\textbf{Theorem 1.} Let $c(N)>0$ be a size-dependent cost function of forming an edge, and let $k$ be a positive integer number specifying the degree of a node in the network. If a network size $\tilde N$ exists starting from which (i.e. for all sizes $N> \tilde N$) the relationship
$$\left[ \frac{2}{f(2) - f(6)} ( c(N) + f(2) - f(6) ) \right] (k^2 +1) < N$$
is satisfied, then the Nash equilibrium will always contain at least a vertex of degree at least $k$. \\

\textbf{Proof.} Let's demonstrate Theorem 1 by contradiction, and  assume the opposite i.e., let us suppose that there is a $k$ such that for any $N$ there are equilibrium states that do not contain a vertex of degree at least $k$.

Due to Lemma 1, this entails the existence of a 2-independent set $S$ of size at least $\frac{N}{k^2+1}$.

Then, due to Lemma 2, any vertex of such a 2-independent set  will increase the value of the benefit function (by forming connections with all other vertices in $S$) by at least ${\frac{N}{k^2+1} - 1\choose 2}\cdot(f(2) - f(6))$. The costs of forming these links will be instead equal to $ (\frac{N}{k^2 + 1} - 1) \cdot c$.

Then one can examine the utility obtained by a vertex $v \in S$ from the formation of such connections, taking into account that, in order to prevent $v$ from connecting with every other vertex in $S$, the difference between the additional utility received and the cost must be negative. One has\\
\begin{center}
	$\frac{1}{2} \cdot  (\frac{N}{k^2 + 1} - 1) \cdot (\frac{N}{k^2 + 1} - 2) \cdot (f(2) - f(6)) - (\frac{N}{k^2 + 1} - 1) \cdot c < 0$ \qquad,
\end{center}
and, therefore,
\begin{center}
	$(\frac{N}{k^2 + 1} - 2) < \frac{2c}{f(2) - f(6)}  \Leftrightarrow \frac{N}{k^2 + 1} < \frac{2c}{f(2) - f(6)} + 2$ \qquad.
\end{center}
In order to be at the Nash equilibrium, it is necessary that the above condition is verified, because otherwise it would be strictly advantageous for a vertex from the maximum 2-independent set $S$ to change its strategy and glue the set $S$. Then, one obtains:
\begin{center}
	$\frac{N}{k^2 + 1} < \frac{2c}{f(2) - f(6)} + 2 \Leftrightarrow N < (\frac{2c}{f(2) - f(6)} + 2)\cdot ({k^2 + 1})  \Longrightarrow N < \left[ \frac{2}{f(2) - f(6)}(c + f(2) - f(6)) \right] (k^2+1)$
\end{center}

which is in direct contradiction with the statement of Theorem 1. This implies that our initial assumption is incorrect, and therefore that there will be necessarily a vertex of degree at least $k$ at the equilibrium ({\it quod erat demonstrandum!}).\\

\subsubsection{The emergence of ultra-small world states and of the six degrees of separation property}

Finally, we can prove the main result of our study, related with the fact that, if they exist, the Nash equilibria are ultra-small world states featuring the six degrees of separation property. \\

\textbf{Theorem 2.} Let the cost $c(N)$ of forming a link in the network satisfy $$c(N) < \sqrt[3]{\frac{f(2) - f(6)}{16}}\cdot \sqrt[3]{N\cdot(f(2) - f(5))^2} - f(2) + f(6).$$
Then, from a given network size $\tilde N$ on (i.e. for all networks whose size $N$ is larger than $\tilde N$), the distance between two generic vertices $v,u$ of the network does not exceed 6 at the Nash equilibrium.\\

\textbf{Lemma 3.} If $$c(N) < \sqrt[3]{\frac{f(2) - f(6)}{16}}\cdot \sqrt[3]{N\cdot(f(2) - f(5))^2} - f(2) + f(6),$$  then all equilibria contain at least a vertex $u$ such that the distance to it from any other vertex of the network does not exceed 3. \\

\textbf{Proof.} Once again, let us proceed by contradiction, i.e. by trying to prove the opposite. Let us then consider a value of the degree $k = \frac{1}{f(2) - f(5)}\cdot \left( \sqrt[3]{\frac{f(2) - f(6)}{2}}\cdot\sqrt[3]{N\cdot(f(2) - f(5))^2} - f(2) + f(6) \right)$. \\

First, one has to notice that the inequality $k^2+1 < 2 k^2$ is always verified, as far as $k > 1$.
Then, one has that
\begin{center}
	$ \frac{2}{f(2) - f(6)}(c + f(2) - f(6))(k^2 + 1) < \frac{2}{f(2) - f(6)}(c + f(2) - f(6)) \cdot 2k^2 = \frac{4}{f(2) - f(6)}(c + f(2) - f(6))\cdot k^2 = $
\end{center}

\begin{center}
	$=\frac{4}{f(2) - f(6)}(c + f(2) - f(6)) \left( \frac{1}{f(2) - f(5)}\cdot(\sqrt[3]{\frac{f(2) - f(6)}{2}}\cdot\sqrt[3]{N\cdot(f(2) - f(5))^2} - f(2) + f(6)) \right)^2  <$
\end{center}

\begin{center}
	$<\frac{4}{f(2) - f(6)} \left( \sqrt[3]{\frac{f(2) - f(6)}{16}}\sqrt[3]{N\cdot(f(2) - f(5))^2} \right) (\sqrt[3]{\frac{f(2) - f(6)}{2}})^2 \cdot $
\end{center}

\begin{center}
	$\cdot \frac{1}{(f(2) - f(5))^2}$ $\cdot \left( \sqrt[3]{N\cdot(f(2) - f(5))^2} \right)^2 = N \Longrightarrow$
\end{center}

\begin{center}
	$\Longrightarrow \frac{2}{f(2) - f(6)}(c + f(2) - f(6))(k^2 + 1) < N.$
\end{center}

Due to Theorem 1, one has therefore to conclude that there must be a vertex of degree $k$ at the Nash equilibrium.

Notice that, in the derivation of the above expression, we have made use of the two inequalities $c + f(2) -f(6) < \sqrt[3]{\frac{f(2) - f(6)}{16}}\cdot \sqrt[3]{N\cdot(f(2) - f(5))^2}$ (from the statement of Lemma 3) and $-f(2)+f(6) < 0$ (from the fact that $f(x)$ is a strictly decreasing function of its argument).

Let us call such a vertex $u$. Suppose now that a vertex $v\in V$ exists such that the distance between $u$ and $v$ is at least 4, and let us calculate the minimum additional utility that the vertex $u$ would accrue by forming a connection with $v$.

We now denote as $S_u$ the set of neighbors of the vertex $u$, and we consider a generic vertex $s_i \in S_u$. If the shortest path from $v$ to $s_i$ passes through $u$, then the distance from $v$ to $s_i$ must be at least 5 (otherwise it would be possible to get to $u$ from $v$ along the edges of the network with less than 3 moves). Notice that the pair $s_i, v$ contributes to $WBC(u)$ no more than $f(l(s_i,v))$, i.e. no more than $f(5)$ in the present case. After forming the edge $(u,v)$, the distance between the vertices $s_i$ and $v$ will become 2, which implies that the contribution of this pair to $WBC(u)$ will become $f(2)$, i.e. it will increase by at least $(f(2) - f(5))$. This argument is valid for all other vertices of $S$. Then, the total benefit from holding the edge $(u,v)$ will be at least $(f(2) - f(5))\cdot k$. At the same time, the cost of forming the edge $(u,v)$ is equal to $c$. Thus, in order to be at the Nash equilibrium, it is necessary that
\begin{center}
	$\left( f(2) - f(5) \right) \cdot k < c \Leftrightarrow k < \frac{c}{f(2) - f(5)}$ .
\end{center}

On the other hand, one has that
\begin{center}
	$ \frac{c}{f(2) - f(5)} < \frac{1}{f(2) - f(5)}\cdot \left( \sqrt[3]{\frac{f(2) - f(6)}{16}}\cdot\sqrt[3]{N\cdot(f(2) - f(5))^2} - f(2) + f(6) \right) < $
\end{center}

\begin{center}
	$ < \frac{1}{f(2) - f(5)}\cdot \left( \sqrt[3]{\frac{f(2) - f(6)}{2}}\cdot\sqrt[3]{N\cdot(f(2) - f(5))^2} - f(2) + f(6) \right) = k$.
\end{center}

Therefore, one has that $ k > \frac{c}{f(2) - f(5)}$, which evidently leads to a contradiction. Therefore, all equilibrium states must necessarily contain at least a vertex $u$ of degree at least $k$ such that each vertex $v\in V$ will necessarily be at a distance of no more than 3 from $u$. Lemma 3 is proved. \\

\textbf{Proof of Theorem 2.} According to Theorem 1, there is a vertex $u$ of degree at least $k$ in the equilibrium state. According to Lemma 3, the distance from any other vertex of the network to $u$ does not exceed 3. This automatically implies that the maximal distance between any two vertices $s,t$  (the diameter of the network) cannot exceed 6 at the equilibrium: no more than 3 steps are needed for passing from $s$ to $u$ and no more than 3 other steps are needed from $u$ to $t$ ({\it quod erat demonstrandum!}). \\

\section{The illustrative case}

In the second part of the main text, we reported on an illustrative model, where nodes can modify their neighborhood only by adding new links to the already existing ones.
In particular, the game consisted in the following steps.

\begin{itemize}
	\item [1)] The agents start the game when they are already connected by means of a pristine graph where, in addition,
	there exists at least one node with sufficiently high degree.
	\item [2)] Each agent adopts as benefit function
	\begin{equation}
		WBC(v)=\sum_{s\ne v\ne t}\frac{\sigma_{st}(v)}{\sigma_{st}}\cdot \frac{1}{l(s,t)^\alpha},	
		\label{benefit2SM}
	\end{equation}
	with $\alpha$ being a strictly positive parameter. This implies that the weighting factor is $f(l(s,t))=\frac{1}{l(s,t)^\alpha}$,
	and that Eq.~(\ref{benefit2SM}) coincides, for $\alpha=1$, with the classical weighted betweenness centrality.
	\item [3)] Agents add new connections to their
	neighborhood if and only if there is a positive balance between the extra-utility brought by the new connections and the extra-cost.
\end{itemize}

In practice, at each step $m$ of the game, each agent
$v \in V$ considers a neighborhood ${\cal N}_v (m)$ which is equal to ${\cal N}_v (m-1)$ plus $p$ other nodes.

When all agents become unable to incorporate further edges, the network is said to have reached its asymptotic equilibrium state.

Under this conditions, we are able to prove the following Theorem: \\

\textbf{Theorem 3.}
If in the initial graph there is a node whose degree $k$ has a value satisfying the relationship
\begin{equation}
	\label{3SM}
	\left(\frac{1}{2^\alpha}-\frac{1}{(H+2)^\alpha}\right)k > c,
\end{equation}
for some integer $H \in \left\{3,4,5,...\right\}$, then the equilibrium state contains at least a node $v$
which is linked to all other nodes of the graph by no more than $H$ links.
This automatically implies that the diameter of the equilibrium state of the network does not exceed $2 H$
and that, therefore, such an equilibrium  state is an ultra-small world state
(and, for $H=3$, also the emergence of the 6 degrees of separation property). \\

\textbf{Proof of Theorem 3.}
The proof is given again by contradiction. i.e. by supposing that there is a node $u$ in the final state of the network whose distance from $v$ is at least $H+1$ i.e., $H'=l(u,v)\geq H+1$. Then, the nodes directly connected to $v$ (the neighbors of $v$) may be found at either $H'-1$, or $H'$, or $H'+1$ edges from $u$.
In the main text, we already discussed that the only contribution  to the benefit function of $v$
from shortest paths ending in $u$ and originating in a neighbor of $v$ is coming from those paths starting in the neighbors of $v$ that are at distance
$H'+1$ from $u$.
The contribution of these paths to Eq.~{\eqref{benefit2SM}} satisfies
\begin{equation*}
	\frac{1}{l^{\;\alpha}_{s,u}} \cdot \frac{\sigma_{s,u}(v)}{\sigma_{s,u}}\leq \frac {1}{(H'+1)^{\;\alpha}}\leq \frac {1}{(H+2)^{\;\alpha}}.
\end{equation*}
This is because $\sigma_{s,u}(v)\leq\sigma_{s,u}$, and $H+2\leq H'+1\leq l_{s,u}$.  Therefore, it easily follows that
\begin{equation*}
	WBC(v)\leq \frac{k}{(H+2)^{\;\alpha}}+R,
\end{equation*}
where $R$ accounts for the contribution of all other shortest paths in the network that pass through $v$ and that have not been considered so far.

When a direct link between nodes $v$ and $u$ is added, the shortest path between any neighbor of $v$ (denoted now generically by $w$) and $u$ becomes $w-v-u$. Consequently,
\begin{equation*}
	\Delta WBC(v)\geq \left(\frac{1}{2^\alpha}-\frac{1}{(H+2)^\alpha}\right) \cdot k,
\end{equation*}
because the contribution of the rest of the shortest paths (which do not start in $w$ and end in $u$) does not decrease (compared to $R$) with the addition of the new edge (see details of the demonstration in the above Lemma 0).
Once again,  let us now recall that the definition of the equilibrium state foresees explicitly that the gain from adding any edge must be smaller than the cost, that is,
\begin{equation*}
	c> \Delta WBC(v)\geq\left(\frac{1}{2^\alpha}-\frac{1}{(H+2)^\alpha}\right) \cdot k,
\end{equation*}
which is in explicit contradiction with condition~(\ref{3SM}) ({\it quod erat demonstrandum!}). \\

\end{document}